\documentstyle[preprint,pre,aps]{revtex}

\begin{document}
\draft
\title{Experimental study of bifurcations in a parametrically forced
       magnetic pendulum
      }

\author{
        Sang-Yoon Kim,
       \footnote{Electronic address: sykim@cc.kangwon.ac.kr}
        Seung-Ho Shin, Jaichul Yi, and Chi-Woong Jang
       }
\address{
Department of Physics, Kangwon National University,
Chunchon, Kangwon-Do 200-701, Korea
}
\maketitle

\begin{abstract}
An experimental study of bifurcations associated with stability of
stationary points (SP's) in a parametrically forced magnetic pendulum
and a comparison of its results with numerical results are
presented. The critical values for which the SP's lose or gain their
stability are experimentally measured by varying the two parameters
$\Omega$ (the normalized natural frequency) and $A$ (the normalized
driving amplitude). It is observed that, when the amplitude $A$
exceeds a critical value, the normal SP with $\theta=0$ ($\theta$ is
the angle between the permanent magnet and the magnetic field) becomes
unstable either by a period-doubling bifurcation or by a
symmetry-breaking pitchfork bifurcation, depending on the values of
$\Omega$. However, in contrast with the normal SP the inverted
SP with $\theta=\pi$ is observed to become stable as $A$ is increased
above a critical value by a pitchfork bifurcation, but it also
destabilizes for a higher critical value of $A$ by a period-doubling
bifurcation. All of these experimental results agree well with
numerical results obtained using the Floquet theory.
\end{abstract}


\narrowtext

\section{Introduction}
\label{sec:Int}

We consider a permanent magnet of dipole moment $m$ placed in a
spatially uniform magnetic field $B$ \cite{Croquette,Schmidt,Briggs}.
Its motion can be described by a second-order nonautonomous ordinary
differential equation,
\begin{equation}
I {\ddot \theta} + b {\dot  \theta} + m (B_{DC}+B_{AC} \sin \omega t)
\sin \theta =0,
\label{eq:MP1}
\end{equation}
where the overdot denotes the differentiation with respect to time,
$\theta$ is the angle between the magnet and the magnetic field, $I$
is the momemt of inertia about a rotation axis, $b$ is the damping
parameter, $B_{DC}$ is the steady DC componet of $B$, and $B_{AC}$ and
$\omega$ are the amplitude and frequency of the sinusoidally
time-varying AC component of $B$, respectively. Making the
normalization
$\omega t \rightarrow 2 \pi t$ and $ \theta \rightarrow 2 \pi x$,
we have
\begin{equation}
 {\ddot x} + 2 \pi \gamma {\dot  x} + 2 \pi
(\Omega^2 + A  \sin 2 \pi t) \sin 2 \pi x =0,
\label{eq:MP2}
\end{equation}
where $\omega_0 = \sqrt{ { {mB_{dc}} \over I}}$,
$\Omega={\omega_0 \over \omega}$, $\gamma={b \over {I \omega}}$,
and $A = {{m B_{ac}} \over {I \omega^2}}$.
Note that this is just a normalized equation of motion for the
parametrically forced gravitational pendulum (GP) with a
vertically oscillating support
\cite{Arnold1,McLaughlin,Leven1,Kim,Leven2,Leven3}.
Hence this magnetic system can be taken as a model of the
parametrically forced pendulum equation.
Hereafter we will call this magnetic oscillator a parametrically
forced magnetic pendulum (MP).

The parametrically forced pendulum, albeit simple looking,
shows a richness in its dynamical behaviors.
As the normalized driving amplitude $A$ is
increased, transitions from periodic attractors to chaotic attracors
and {\it vice} {\it versa}, coexistence of different attractors,
transient chaos, multiple period-doubling transitions to chaos, and so
on have been numerically found \cite{McLaughlin,Leven1,Kim}. Some of
them have also been observed in real experiments on the parametrically
forced MP \cite{Croquette,Schmidt,Briggs} and GP \cite{Leven2,Leven3}.
However, so far only the case of $\Omega=0$  (i.e., the case
$B_{DC}=0$) has been studied in the experiments on the MP.

In this paper, we study experimentally the bifurcations associated
with stability of the SP's in the parametrically forced MP
by varying the normalized natural frequency $\Omega$ and the
normalized driving amplitude $A$, and then compare the experimental
results with the numerical results obtained using the Floquet
theory \cite{Gukenheimer1}. We first note that the MP has two SP's.
One is the ``normal'' SP with $(x,{\dot x})=(0,0)$, and the other one
is the ``inverted'' SP with $(x,{\dot x})=({1 \over 2},0)$.
For the case of the ``unforced'' simple MP (with $A=0$), the normal SP
is stable, while the inverted SP is unstable. However, as $A$ is
increased above a critical value, the normal SP loses its stability.
Using the Floquet theory, bifurcations occurring at such critical
values have been numerically studied by one of us (Kim) and
Lee \cite{Kim}. In contrast to the normal SP, the inverted SP gains
its stability when $A$ exceeds a critical value
\cite{Stoker,Corben,Kapitza,Arnold2,Blackburn}. We also study
numerically bifurcations associated with stability of the inverted SP
using the Floquet theory for a comparison with the experimental
results.

This paper is organized as follows.
We first explain the experimental setup for the parametrically forced
MP in Sec.~\ref{sec:ES}. Bifurcations of the two SP's are then
experimentally investigated in Sec.~\ref{sec:BIF}. It is observed that
as $A$ is increased beyond a critical value, the normal SP loses its
stability either by a period-doubling bifurcation or by a
symmetry-breaking pitchfork bifurcation, depending on the values of
$\Omega$. For the case of the period-doubling bifurcation, a new
stable symmetric orbit with period $2$ is born, while for the case of
the symmetry-breaking pitchfork bifurcation a conjugate pair of new
stable asymmetric orbits with period $1$ appears. In contrast with the
normal SP, the inverted SP is observed to gain its stability when $A$
exceeds a 1st critical value $A^*_1$ by a pitchfork bifurcation, but
it also destabilizes for a higher 2nd critical value $A^*_2$ of $A$ by
a period-doubling bifurcation. Thus the inverted SP becomes stable in
the interval between $A^*_1$ and $A^*_2$. Our experimental data shows
a good agreement between the experimental and numerical results.
Finally, Sec.~\ref{sec:SUM} gives a summary.

\section{Experimental Setup}
\label{sec:ES}

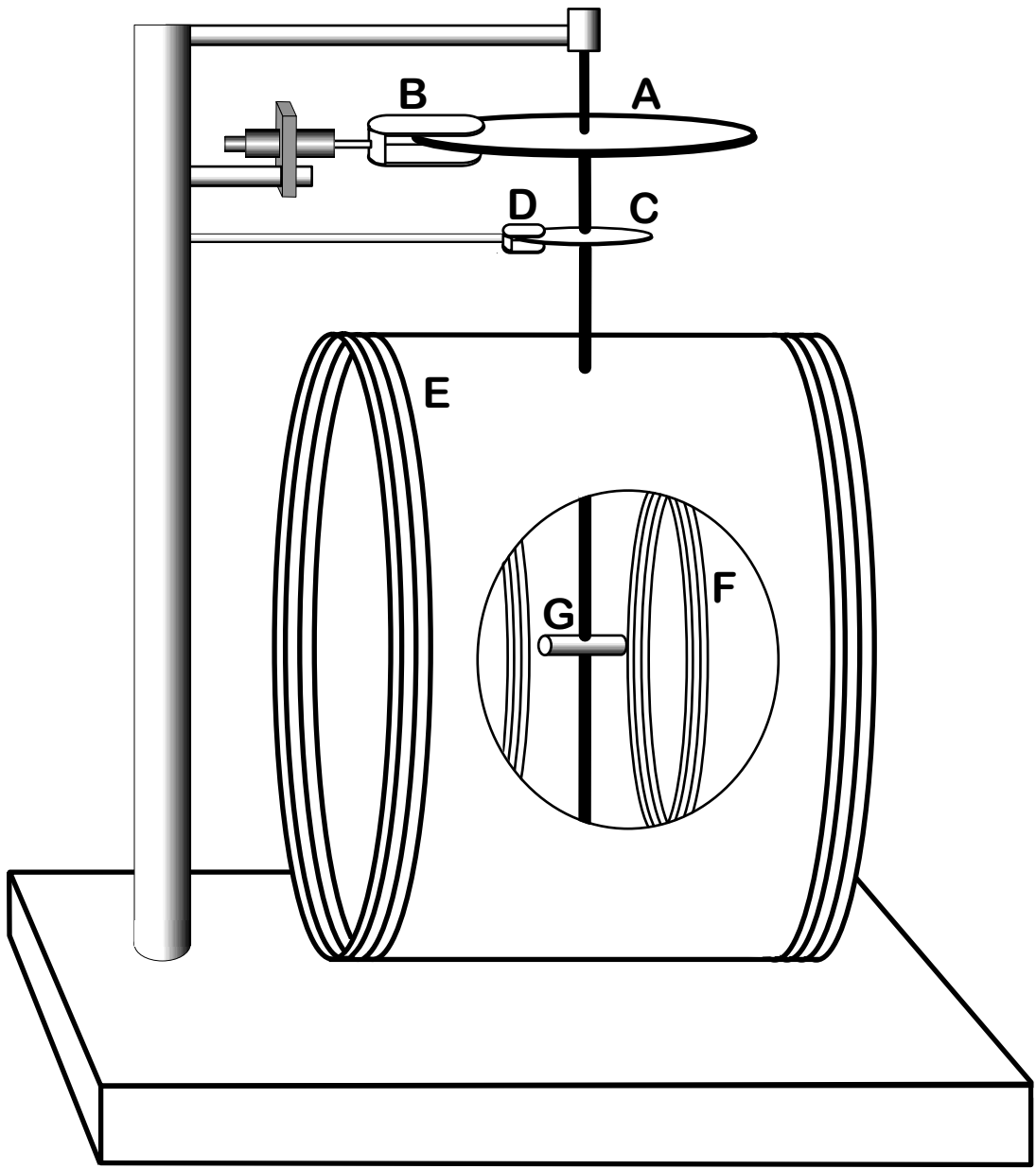
\begin{figure}
\caption{Exploded view of a parametrically forced MP.
         Labeled components are the aluminum damping plate ($A$), the
         horseshoe magnet controlled by a micrometer $(B)$, the code
         wheel $(C)$, the encoder module $(D)$, the Helmholtz coil for
         production of a steady DC component of a spatially uniform
         magnetic field $B$ $(E)$, the Helmholtz coil for production of
         a sinusoidally time-varying AC component of $B$ $(F)$ and the
         permanent bar magnet $(G)$.}
\label{fig:EV}
\end{figure}

An exploded view of the experimental apparatus is shown in
Fig.~\ref{fig:EV}. The physical MP consists of a permanent bar magnet,
an aluminum damping plate, and a code wheel, which are coaxially
attached to a rotation axis. This MP shows rich dynamical behaviors
in a spatially uniform magnetic field $B$, generated by Helmholtz
coils. Each component of the apparatus is explained in some details
below.

A permanent bar magnet, glued to a rotation axis guided by a small
ball in the lowest part and by a tiny pin in the highest part,
is placed in the center of two sets of Helmholtz coils, producing
the magnetic field $B$ perpendicular to the rotation axis. The number
of turns $N$ and the radius $R$ for a large set of Helmholtz coils are
$N=130$ and $R=10.8$ cm, while they are $N=144$ and $R=5.8$ cm
for a smaller set of Helmholtz coils.
The large set of Helmholtz coils is given a direct current $I_{DC}$
to supply a steady DC component $B_{DC}$ of $B$. On the other hand,
the smaller set of Helmholtz coils, which is nested inside the large
set and driven by a Pasco Model CI-6552A AC power amplifier, provides
a sinusoidally time-varying AC component of $B$
with amplitude $B_{AC}$ and frequecny $\omega$. This amplifier can
handle currents up to $1$ A. Since higher currents are necessary in
some range of $\Omega$, we also use a current booster to increase the
current up to $2$ A.

We take into account the effect of the normal component $B_{E,n}$
of the Earth's magnetic field $B_{E}$ perpendicular to the rotation
axis, and align both the DC and AC components of the applied magnetic
field $B$ parallel to $B_{E,n}$.
The angle of the permanent bar magnet is experimentally measured from
this aligned direction (i.e., the direction of $B_{E,n}$).
Since the effective DC component $B_{eff}$ of the total magnetic field
is given by $B_{eff}=B_{DC}+B_{E,n}$, the natural frequency $\omega_0$
in Eq.~(\ref{eq:MP2}) becomes
$\omega_0 = \sqrt{ {m \over I} B_{eff}}$,
where $m/I = 2.542$ gauss${}^{-1}$\,s${}^{-2}$ and $B_{E,n}=0.223$
gauss \cite{BE}.

A damping force proportional to the angular velocity can be supplied
by an eddy current brake. Such eddy-current damping is adjusted by
controlling the separation between the aluminum damping plate and
the horseshoe magnet with a micrometer screw.
It is also possible to determine a damping parameter $k$
$(\equiv b/2I)$ and the natural frequency $\omega_0$ by fitting a
sampled time-series $\theta (t)$ for the unforced case of $B_{AC}=0$
to an equation,
$\theta(t) = C e^{-kt} \cos{(\sqrt{\omega_0^2 - k^2}\,t + \delta)}$
($C$ and $\delta$ are some constants), which is just the angle in the
case of the underdamped motion of the unforced MP for small angular
displacements.

For data acquisition and experimental control, we use commercial
products, ``Rotary Motion Sensor'' and ``Signal Interface,''
manufactured by the Pasco Scientific. A Pasco Model CI-6538 rotary
motion sensor set consists of a code wheel with $1440$ slots (i.e., a
disk with angular code in the form of sectors which are pervious or
impervious to light) and an encoder module containing
a light-emitting diode (LED) and two photodiodes with
signal-processing circuitry.
As the code wheel moves, the light signal emitted from the LED
is interrupted by the slots and electrically encoded.
Thus a code wheel with $1440$ slots can
generate raw data with a resolution of $\Delta \theta=0.004$ rad.
A personal computer equipped with a Pasco Model CI-6560 signal
interface unit analyses the signals from the rotary motion sensor
and provides an easy-to-use data set of $(\theta, {\dot {\theta}})$
at a chosen sampling rate. It also performs a convenient experimental
control.

\section{Bifurcations of the normal and inverted SP's}
\label{sec:BIF}

In this section we first analyse the bifurcations associated with
stability of the SP's in the parametrically
forced MP, using the Floquet theory. The experimental results for the
cases of the normal and inverted SP's are then presented and compared
with the numerical results.

\subsection{Bifurcation analysis based on the Floquet theory}
\label{subsec:BA}

The normalized second-order ordinary differential equation
(\ref{eq:MP2}) is reduced to two first-order ordinary differential
equations:
\begin{mathletters}
\begin{eqnarray}
{\dot x} &=& y,  \\
{\dot y} &=& -2 \pi \gamma y - 2 \pi (\Omega_0^2  + A \sin 2 \pi t)
\sin 2 \pi x.
\end{eqnarray}
\label{eq:MP3}
\end{mathletters}
These equations have an inversion symmetry $S$, since the
transformation
\begin{equation}
S: x \rightarrow -x,\, y \rightarrow -y,\, t \rightarrow t,
\end{equation}
leaves Eq.~(\ref{eq:MP3}) invariant.
If an orbit $z(t)$ $[\equiv (x(t),y(t))]$ is invariant under $S$, it
is called a symmetric orbit. Otherwise, it is called an asymmetric
orbit and has its ``conjugate'' orbit $Sz(t)$ $[=(-x(t),-y(t))]$.

The Poincar\'{e} maps of an initial point $z_0 [=(x(0),y(0))]$ can be
computed by sampling the orbit points $z_m$ at the discrete time
$t=m$, where $m=1,2,3,\dots$~. We call the transformation
$z_m \rightarrow z_{m+1}$
the Poincar\'{e} (time-1) map, and write $z_{m+1} = P (z_m)$.
Note that the normal and inverted SP's of the parametrically forced
MP, denoted by $\hat{z}_N$ $[\equiv (0,0)]$ and $\hat{z}_I$
$[\equiv ({1 \over 2},0)]$, respectively, are the fixed points (FP's)
of the Poincar\'{e} map $P$ [i.e., $P(\hat{z})=\hat{z}$ for
$(\hat{z} = \hat{z}_N, \hat{z}_I)]$
with period $1$.

Here we investigate bifurcations associated with stability of the two
normal and inverted FP's of $P$. The linear stability of an FP is
determined from the linearized-map matrix $DP$ of $P$ at $\hat{z}$.
Using the Floquet theory \cite{Gukenheimer1}, we obtain the matrix
$DP$ by integrating the linearized differential equations for small
perturbations as follows. Consider an infinitesimal perturbation
$[\delta x(t), \delta y(t)]$ to an SP. Linearizing
Eq.~(\ref{eq:MP3}) about the SP, we obtain
\begin{equation}
 \left( \begin{array}{c}
         \delta {\dot x}  \\
         \delta {\dot y}
      \end{array}
      \right)
      = J(t)
      \left( \begin{array}{c}
         \delta x  \\
         \delta y
      \end{array}
      \right),
\label{eq:LEQ}
\end{equation}
where
\begin{equation}
 J(t) =
 \left( \begin{array} {cc}
        0 & 1 \\
        -4 \pi^2 (\Omega^2 + A \sin 2 \pi t) \cos 2 \pi \hat{x} &
        -2 \pi \gamma
        \end{array}
 \right).
\end{equation}
Here $\hat{x} =0$ and $1 \over 2$ for the normal and inverted SP's,
respectively.

Note that $J$ is a $2 \times 2$ $1$-periodic matrix
[i.e., $J(t+1) = J(t)$]. Let $W(t)=[w^1(t),w^2(t)]$ be a fundamental
solution matrix  with $W(0) = I$. Here $w^1(t)$ and $w^2(t)$ are two
independent solutions expressed in column vector forms, and $I$ is the
$2 \times 2$ unit matrix. Then a general solution of the $1$-periodic
system has the following form
\begin{equation}
 \left( \begin{array}{c}
         \delta x(t)  \\
         \delta y(t)
      \end{array}
      \right)
      = W(t)
      \left( \begin{array}{c}
         \delta x(0)  \\
         \delta y(0)
      \end{array}
      \right).
\label{eq:FSM}
\end{equation}
Substitution of Eq.~(\ref{eq:FSM}) into Eq.~(\ref{eq:LEQ}) leads to
an initial-value problem in determining $W(t)$:
\begin{equation}
{\dot W(t)} = J(t) W(t), \;\;W(0)=I.
\label{eq:FSMEQ}
\end{equation}
It is  clear   from  Eq.~(\ref{eq:FSM})  that  $W(1)$   is  just  the
linearized-map matrix   $DP$.   Hence   the  matrix   $DP$
can be  calculated through integration of Eq.~(\ref{eq:FSMEQ}) over
the period $1$.

The characteristic equation of the linearized-map matrix
$M (\equiv DP)$ is
\begin{equation}
\lambda^2 - {\rm tr}M \, \lambda + {\rm det} \, M = 0,
\end{equation}
where ${\rm tr}M$ and ${\rm det}M$ denote the trace and determinant of 
$M$, respectively.   The eigenvalues,   $\lambda_1$   and $\lambda_2$,
of $M$ are called the Floquet (stability) multipliers of the FP. As
shown in \cite{Lefschetz}, ${\rm det}\,M$ is calculated from a formula
\begin{equation}
{\rm det}\,M = e^{\int_0^1 {\rm tr}\,J dt}.
\label{eq:Det}
\end{equation}
Substituting the trace  of $J$ (i.e., ${\rm  tr} J = -  2 \pi \gamma$)
into Eq.~(\ref{eq:Det}), we obtain
\begin{equation}
{\rm det}\,M = e^{-2 \pi \gamma }.
\end{equation}
Note that $M$ is just a $ 2 \times 2$ matrix with a constant Jacobian
determinant (less than unity). Hence the pair of Floquet multipliers
of an FP lies either on the circle  of radius  $e^{-\pi \gamma}$ or
on the  real axis
in the complex plane. The FP is stable  only when both Floquet
multipliers lie inside the unit circle. We first note that they never
cross the unit circle except at the real axis (i.e., they never have
complex values with moduli larger than unity), and hence Hopf
bifurcations do not occur. Consequently, the FP can lose its stability
only when a Floquet multiplier $\lambda$ decreases (increases)
through $-1$ $(1)$ on the real axis.

When a Floquet multiplier $\lambda$ decreases through $-1$, the FP
loses its stability via period-doubling bifurcation, which leads to
the birth of a new stable symmetric orbit with period $2$. On
the other hand, when a Floquet multiplier $\lambda$ increases through
$1$, it becomes unstable via pitchfork bifurcation, which results in
the birth of a conjugate pair of new stable asymmetric orbits with
period $1$. Since the newly-born orbits are asymmetric ones, the
pitchfork bifurcation is also called a symmetry-breaking bifurcation.
For more details on bifurcations, refer to Ref.~\cite{Gukenheimer2}.

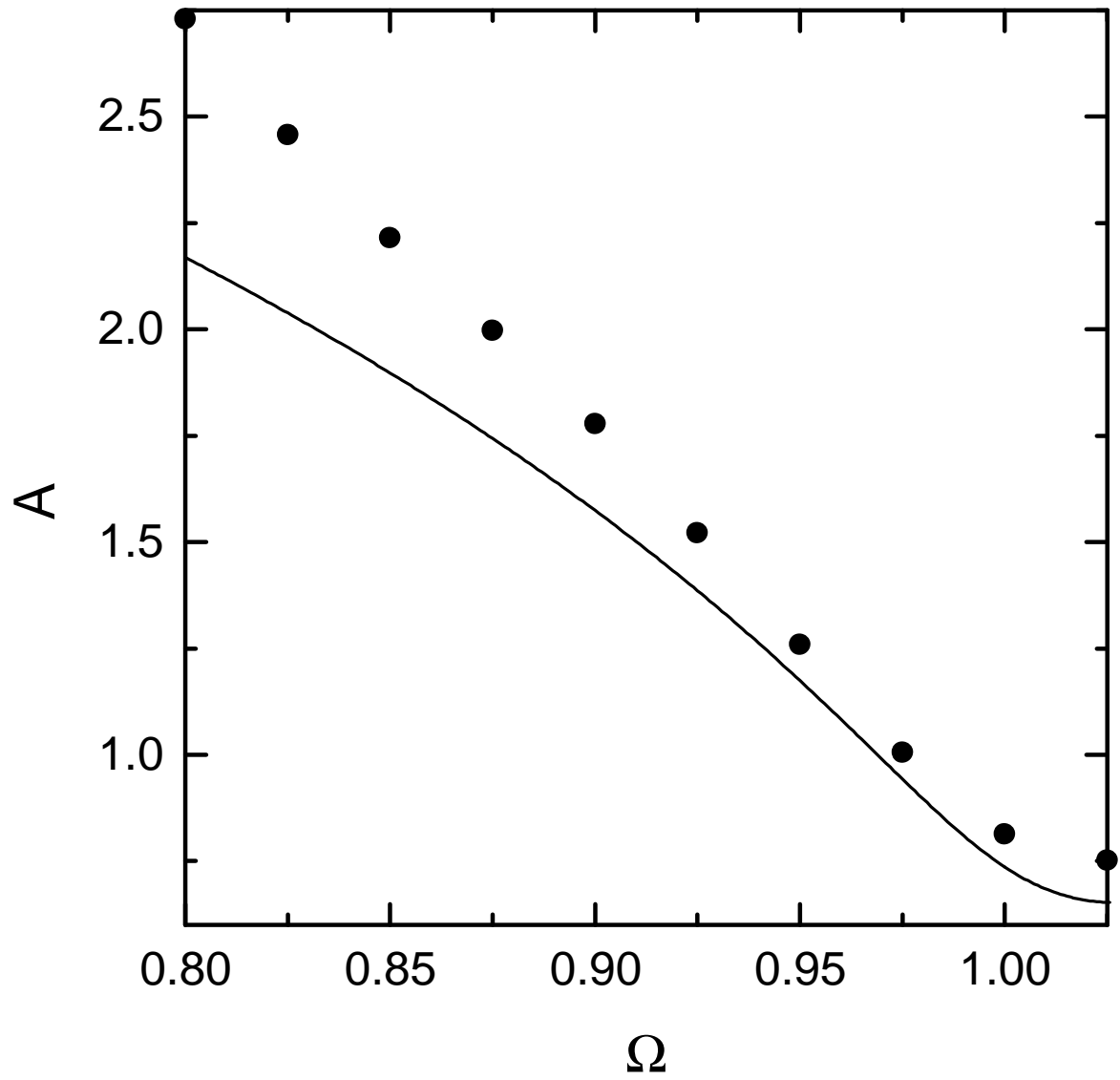
\begin{figure}
\caption{Stability  diagram of  the normal SP in the 2nd range $II$ of
         $\Omega$ (i.e., $0.8 \leq \Omega \leq 1.025$). The
         experimental data for the critical values $A^*_{exp}$'s,
         above which the SP becomes unstable, are represented by the
         solid circles. The stability boundary numerically determined
         using the Floquet theory is denoted by the solid line, and
         it is just a symmetry-breaking pitchfork bifurcation line.
         }
\label{fig:NSD2}
\end{figure}

The stability boundaries of the normal and inverted SP's in some
ranges of the $\Omega$-$A$ plane are determined through numerical
calculations of their Floquet multipliers $\lambda$'s. The absolute
value of $\lambda$ at such stability boundaries is one (i.e.,
$|\lambda|=1$). If $\lambda=-1$, then the boundary is a
period-doubling bifurcation line. Otherwise, it is a
symmetry-breaking pitchfork bifurcation line. We also obtain
numerically the bifurcation diagrams and the phase-flow and
Poincar\'{e}-map plots at some chosen parameter values for clear
visual representation of the bifurcations. All of these numerical
results are given in the next two subsections
(see Figs.~\ref{fig:NBD1}-\ref{fig:ISD})
for a comparison with the experimental results.

\subsection{Experimental results for the case of the normal SP}
\label{subsec:NSP}

In all the experiments for the normal and inverted SP's, we fix the
the driving frequency $\omega$ in Eq.~(\ref{eq:MP1}) and the
normalized damping parameter $\gamma$ in Eq.~(\ref{eq:MP2})
as $\omega=2 \pi$ and $\gamma=0.1$, respectively.
We then control the normalized natural frequency and driving
amplitude, $\Omega$ and $A$, in Eq.~(\ref{eq:MP2}) by varying $B_{DC}$
and $B_{AC}$, respectively and study the bifurcations of the two SP's.

We consider two ranges of $\Omega$ for the normal SP,
$I: 0.2 \leq \Omega \leq 0.5$ and $II: 0.8 \leq \Omega \leq 1.025$.
For each chosen value of $\Omega$, we increase the amplitude $A$ and
observe whether the SP is stable or not. In order to experimentally
determine the stability of the SP, we release the MP from rest at a
small initial angle displaced from the SP. If the SP is stable, then
the subsequent motion damps toward the SP. Otherwise, it deviates from
the SP. Thus we measure experimentally a critical value $A^*_{exp}$
of $A$, above which the SP is unstable.

We first study the bifurcations of the normal SP in the 1st range $I$
of $\Omega$. As an example, consider the case of $\Omega=0.4$.
With increasing $A$, we carry out the experiments, and measure the
critical value $A^*_{exp}$. It is observed that for
$A > A^*_{exp}$, the SP becomes unstable through a period-doubling
bifurcation, giving rise to the birth of a new stable symmetric
period-doubled orbit. For visual representation of the bifurcation, we
obtain the bifurcation diagram and the phase-flow and Poincar\'{e}-map
plots below.

For phase representation, we acquire a data set of
$[\theta(t), {\dot {\theta}}(t)]$ at a fixed sampling rate $20$ Hz
and convert it into a normalized set of $(x,y)$.
This whole set of the data is used for a phase-flow plot, while a
partial set of the data chosen at integral multiples of the external
driving period $T$ $(=2 \pi)$ [i.e., $t=n T\, (n=0,1,2,...)]$ is used
for a Poincar\'{e}-map plot and for a bifurcation diagram.

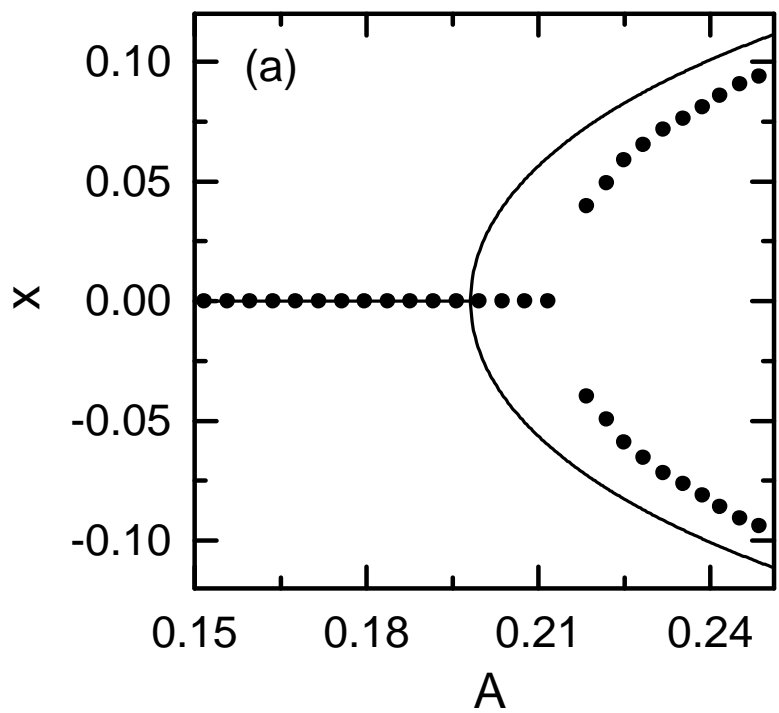
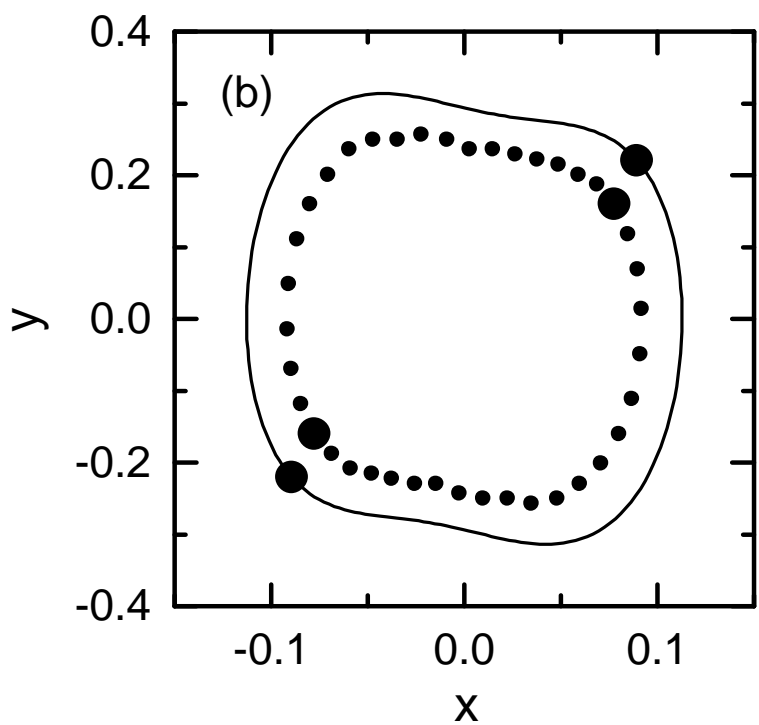
\begin{figure}
\caption{Period-doubling bifurcation of the normal SP for
         $\Omega=0.4$. As shown in the bifurcation diagram (a), the
         normal SP denoted by the solid circles
         becomes unstable for a critical value $A^*_{exp}$ $(=0.215)$
         by a period-doubling bifurcation, which results in
         the birth of a symmetric period-doubled orbit denoted by the
         solid circles. The experimental data for the phase flow for
         $A=0.23$ are denoted by the small solid circles in (b).
         Numerical data denoted by the solid lines are also given in
         both (a) and (b). The data for the Poincar\'{e} maps
         in (b) are represented by the two large solid circles for
         both the experimental and numerical cases. For more details
         see the text.
         }
\label{fig:NBD1}
\end{figure}

The bifurcation diagram for $\Omega=0.4$ is shown in
Fig.~\ref{fig:NBD1}(a).
The data obtained through numerical calculations are also given for
a comaprison with the experimental results. Note that for the
bifurcation diagram, the experimental data represented by the
solid circles agree well with the numerical data denoted by the
solid lines. For reference, the critical values obtained through
experiments and numerical calculations are $A^*_{exp}=0.215$ and
$A^*_{nu}=0.198\,131 \cdots$, respectively. As $A$ is increased
above the critical value $A^*$, the normal SP loses its stability
via period-doubling bifurcation, and a new stable symmetric
period-doubled orbit appears. Figure \ref{fig:NBD1}(b) shows the
phase-flow and Poincar\'{e}-map plots of the symmetric period-doubled
orbit for $A=0.23$. The experimental data for the
phase flow are represented by the small solid circles, while the two
larger solid circles denote the experimental data for the
Poincar\'{e} map. These experimental data are also in a good agreement
with the numerically-computed data for the phase flow represented by
the solid line and for the Poincar\'{e} map denoted by the two large
circles.

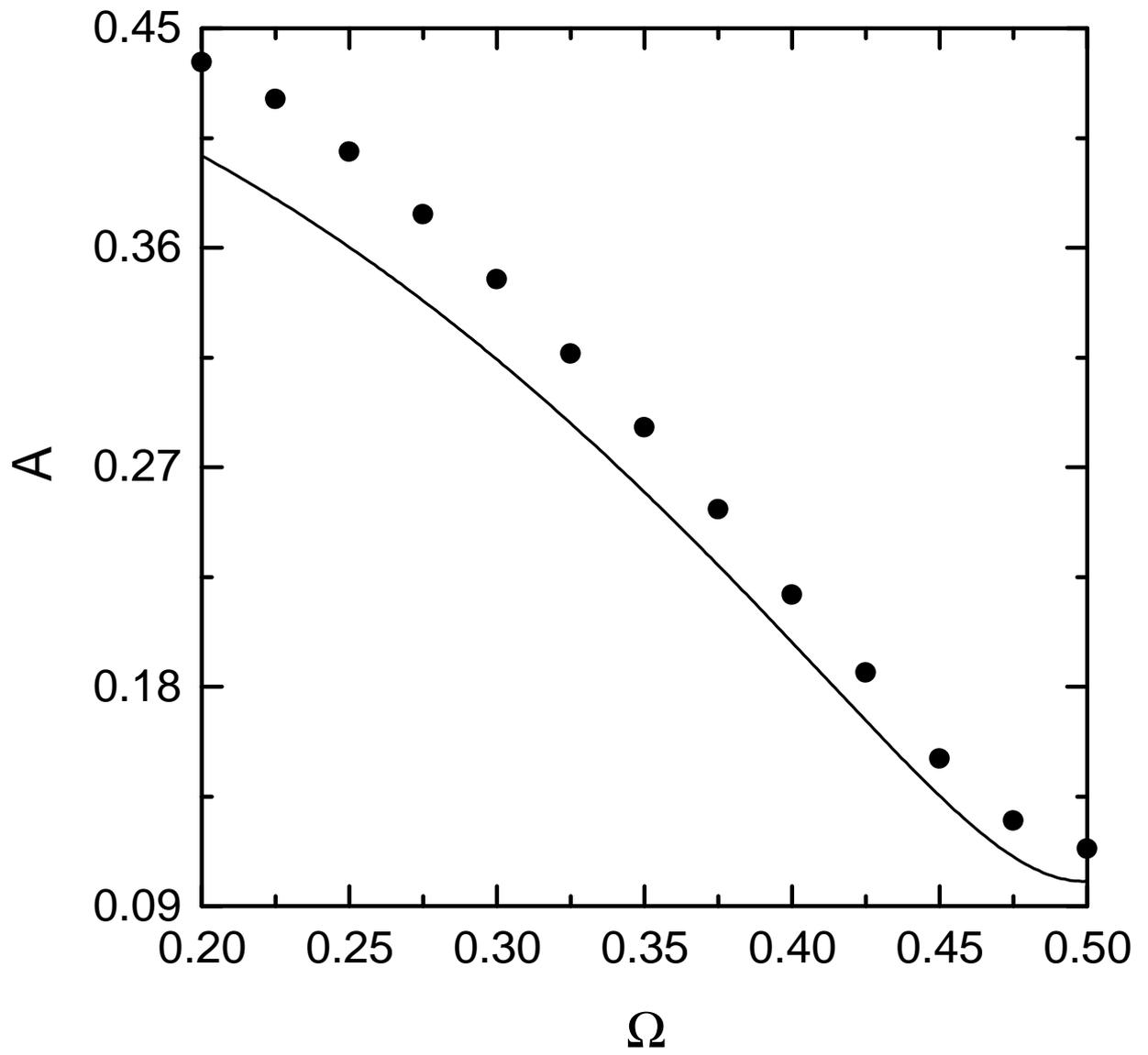
\begin{figure}
\caption{Stability  diagram of  the normal SP in the 1st range $I$ of
         $\Omega$ (i.e., $0.2 \leq \Omega \leq 0.5$). The experimental
         data for the critical values $A^*_{exp}$'s, above which the
         SP becomes unstable, are represented by the solid circles.
         The stability boundary numerically determined using the
         Floquet theory is denoted by the solid line, and it is just
         a period-doubling bifurcation line.
     }
\label{fig:NSD1}
\end{figure}

We also perform the above experiments for many other values of
$\Omega$ in the 1st range $I$ and thus measure the critical values
$A^*_{exp}$'s. The stability diagram of the normal SP is shown in
Fig.~\ref{fig:NSD1}. The experimental data for $A^*_{exp}$ are
represented by the solid
circles, and they seem to lie on a smooth stability boundary curve.
For a comparison with the experimental data, the stability boundary
of the SP numerically calculated using the Floquet theory is also
denoted by the solid line in Fig.~\ref{fig:NSD1}. This stability
boundary is just a period-doubling bifurcation line at which a
Floquet multiplier of the SP is $\lambda=-1$. The period-doubling
bifurcation line determined through numerical computations lies a
little below that experimentally determined. That is, the value of
$A^*_{exp}$ is somewhat higher than that of $A^*_{nu}$. This is what
one would expect, because in real experiments there exists a
frictional force  due to a contact between the rotation axis and its
guiders (ball and pin).
As previously noted \cite{Leven2,Leven3}, one of the main effects of
this frictional force is to make the origin of the phase plane (i.e.,
the normal SP) stable up to higher values of the external driving
amplitude than the numerically-calculated critical value $A^*_{nu}$.

We now study the bifurcations of the normal SP in the 2nd range $II$
of $\Omega$ (i.e., $0.8 \leq \Omega \leq 1.025$). As in the above
1st range $I$ of $\Omega$, we increase the amplitude $A$ and measure
a critical value $A^*_{exp}$, beyond which the SP becomes unstable,
by releasing the MP from rest at a small initial angle displaced
from the SP. However, in contrast to the 1st range of
$\Omega$ the normal SP is observed to lose its stability through a
symmetry-breaking pitchfork bifurcation for $A=A^*_{exp}$, which leads
to the birth of a conjugate pair of new stable asymmetric orbits
with period $1$.

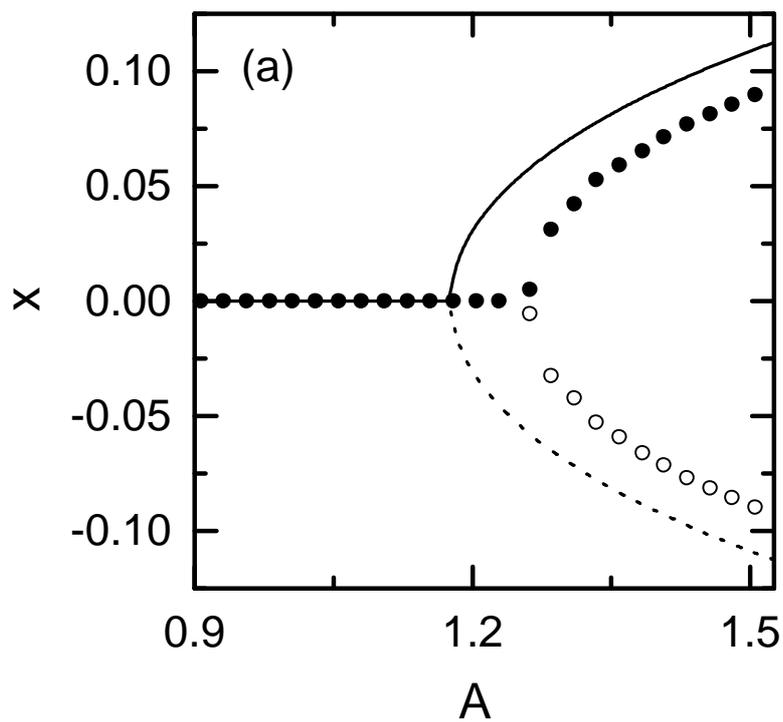
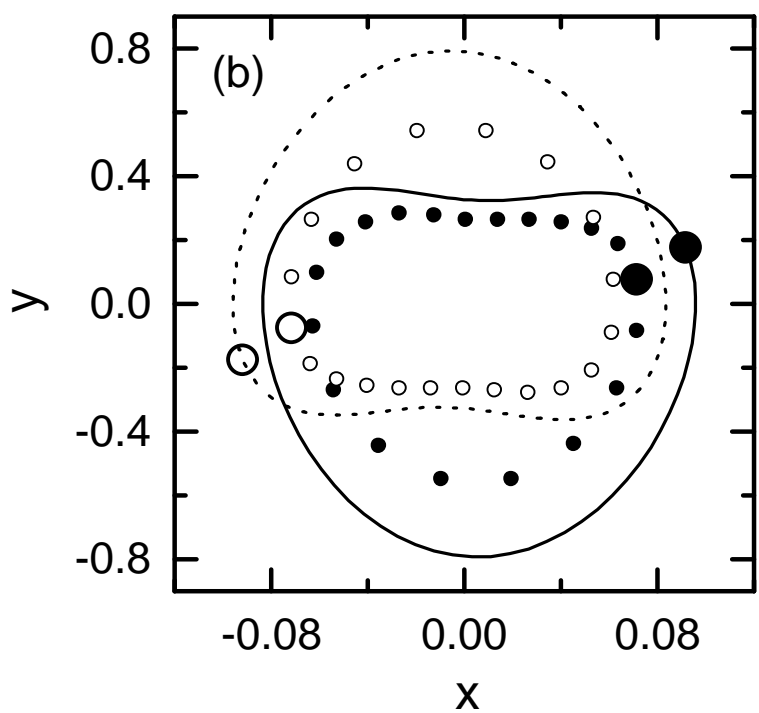
\begin{figure}
\caption{Symmetry-breaking pitchfork bifurcation of the normal SP for
         $\Omega=0.95$. For a critical value $A^*_{exp}$ $(=1.25)$,
         the normal SP denoted by the solid circles
         becomes unstable by a symmetry-breaking
         pitchfork bifurcation, as shown in the bifurcation diagram
         (a). As a result of the bifurcation,
         a conjugate pair of asymmetric orbits of period $1$ appears;
         one is denoted by the solid circles, while the other one is
         represented by the open circles. The experimental data for
         the phase flow of the two symmetry-broken orbits for $A=1.4$
         are denoted by the small solid and open circles in (b),
         respectively. Numerical data for the two symmetry-broken
         orbits are also given in both (a) and (b); one is denoted by
         the solid line, while the other one is represented by the
         dotted line. The data for the Poincar\'{e} maps in (b) are
         represented by the large solid and open circles for both
         the experimental and numerical cases. For more details see
         the text.
         }
\label{fig:NBD2}
\end{figure}

As an example, consider the case of $\Omega=0.95$. The bifurcation
diagram for this case is shown in Fig.~\ref{fig:NBD2}(a). The normal
SP denoted by the solid circles is observed to become  unstable
through a symmetry-breaking pitchfork bifurcation for
$A^*_{exp} =1.25$. For $A>A^*_{exp}$, a pair of stable asymmetric
orbits of period $1$ appears. One is represented by the solid
circles, while its conjugate orbit is denoted by the open circles.
Figure \ref{fig:NBD2}(b) shows the phase-flow and Poincar\'{e}-map
plots of a conjugate pair of symmetry-broken orbits of period $1$ for
$A=1.4$. The small solid circles denote the phase flow of an
asymmetric ``heart-shaped'' orbit, while the small open circles
represent the phase flow of its conjugate ``inverted heart-shaped''
orbit. The data for the Poincar\'{e} maps of the two symmetry-broken
orbits are also denoted by the larger solid and open circles,
respectively. This symmery-broken case is in contrast to the
symmetry-preserved case [see Fig.~\ref{fig:NBD1}(b)] in the 1st range
of $\Omega$. For a comparison with the experimental results, the data
obtained by numerical computations are also given in
Fig.~\ref{fig:NBD2}. As $A$ is increased above a critical value
$A^*_{nu}$ $(=1.174\,209 \cdots)$, the normal SP denoted by the solid
line becomes unstable through a
symmetry-breaking pitchfork bifurcation, giving rise to the birth of
a conjugate pair of symmetry-broken orbits of period $1$. One
asymmetric orbit is represented by the solid line, while the other one
is denoted by the dotted line. As in the experimental case, the data
for the Poincar\'{e} maps of the two asymmetric orbits are denoted by
the large solid and open circles, respectively.
All of these experimental and numerical results seem to agree well.

We also measure the critical values $A^*_{exp}$'s for many other
values of $\Omega$ in the 2nd range $II$. Figure \ref{fig:NSD2} shows
the stability diagram for this case. The experimental data for
$A^*_{exp}$ are denoted by the solid circles, while the stability
boundary numerically computed using the Floquet theory is represented
by the solid line. In contrast to the 1st
range of $\Omega$, the stability boundary is a symmetry-breaking
pitchfork bifurcation line at which a Floquet multiplier of the SP
is $\lambda=1$. We also note that as in the case of the 1st range $I$,
the value of $A^*_{exp}$ is somewhat higher than that of $A^*_{nu}$
because of the frictional force between the rotation axis and its
guiders.

\subsection{Experimental results for the case of the inverted SP}
\label{subsec:ISP}

In this subsection, we study the bifurcations associated with
stability of the inverted SP by increasing $A$ in a range of
$0.2 \leq \Omega \leq 0.5$. In contrast to the normal SP, the inverted
SP is observed to gain its stability when a 1st critical value $A^*_1$
of $A$ is
exceeded by a subcritical pitchfork bifurcation. However, as $A$ is
further increased, the stabilized inverted SP is also observed to lose
its stability for a 2nd critical value $A^*_2$ of $A$ through a
period-doubling bifurcation. Thus the inverted SP becomes stable in
the interval between $A^*_1$ and $A^*_2$.

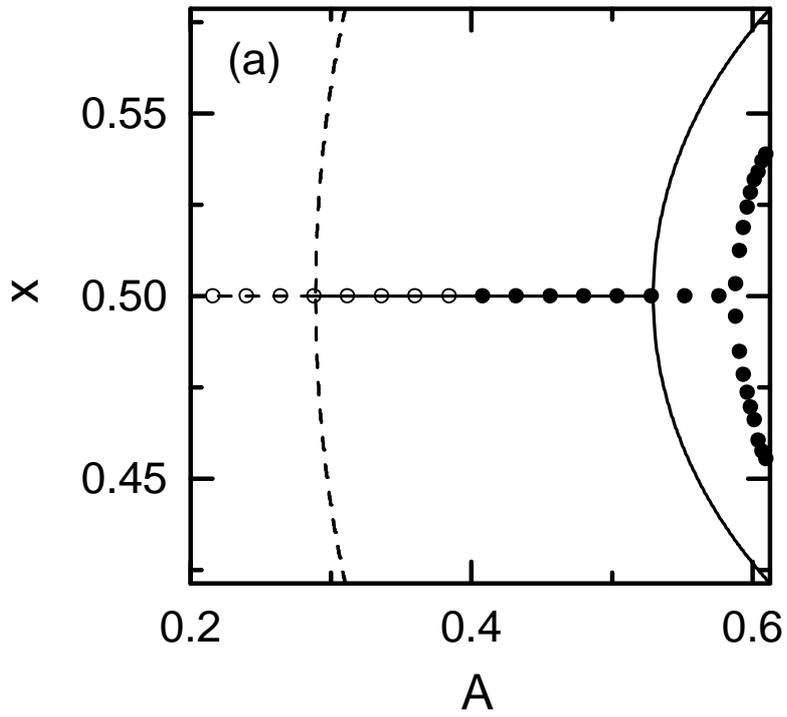
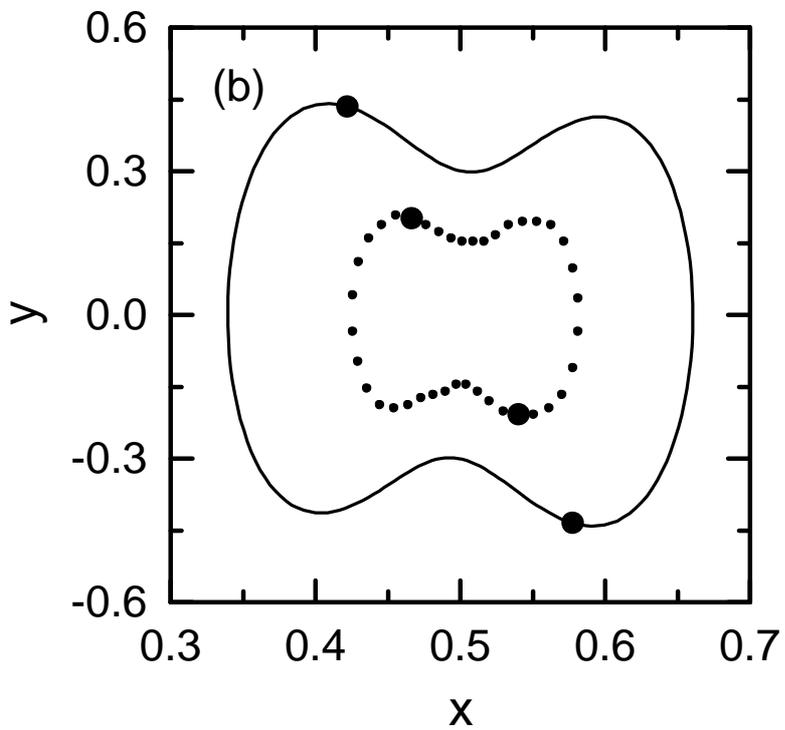
\begin{figure}
\caption{Bifurcations of the inverted SP for $\Omega=0.2$. The
         bifurcation diagram for $\Omega=0.2$ is shown in (a). In
         contrast to the normal SP, the inverted SP denoted by the
         open circles is observed to become stable as $A$ is increased
         above a 1st critical value $A^*_{exp,1}$ $(=0.39)$. However,
         as $A$ is further increased, the stabilized inverted SP
         denoted by the solid circles is also observed to lose its
         stability for a 2nd critical
         value $A^*_{exp,2}$ $(=0.567)$ through a period-doubling
         bifurcation, giving rise to the birth of a symmetric
         period-doubled orbit denoted by the solid circles. The
         experimental data for the phase flow of the period-doubled
         orbit for $A=0.61$ are denoted by the small solid circles in
         (b). Numerical data
         denoted by the lines are also given in both (a) and (b);
         a stable orbit is denoted by a solid line, while an unstable
         orbit is represented by a dashed line.
         The data for the Poincar\'{e} maps in (b) are represented by
         the two large solid circles for both the experimental and
         numerical cases. For more details see the text.
         }
\label{fig:IBD}
\end{figure}

As an example, we consider the case of $\Omega=0.2$. The bifurcation
diagram for this case is shown in Fig.~\ref{fig:IBD}(a). The
unstable inverted SP denoted by the open circles is observed to
become stable when $A$ is increased above a 1st
critical value $A^*_{exp,1}$ $(=0.39)$. Using the Floquet theory, the
unstable inverted SP denoted by the dashed line is also numerically
found to gain its stability for $A> A^*_{nu,1}$ $(=0.289\,108 \cdots)$
by a subcritical pitchfork bifurcation, giving rise to the birth of a
conjugate pair of unstable asymmetric orbits with period $1$, denoted
by the dashed lines. However, unfortunately the two symmetry-broken
orbits born for this subcritical case cannot be experimentally
observed, because they are unstable ones.
This is in contrast to the supercritical bifurcations occurring for
the normal SP in the 2nd range $II$ of $\Omega$ (for a supercritical
case, a pair of stable asymmetric orbits is born, which can be
experimentally observed as shown in Fig.~\ref{fig:NBD2}). As $A$ is
further increased from $A^*_{exp,1}$, the stabilized inverted SP
denoted by the solid circles is observed to lose its stability by a
period-doubling bifurcation when a
second critical value $A^*_{exp,2}$ $(=0.567)$ is exceeded. For
$A > A^*_{exp,2}$, a stable symmetric ``butterfly-shaped''
orbit of period $2$ appears. Small solid circles and the two larger
solid circles in Fig.~\ref{fig:IBD}(b) represent the phase flow and
Poincar\'{e} map of the symmetric orbit of period $2$ for $A=0.61$,
respectively. It is also numerically found that the stabilized
inverted SP denoted by the solid line becomes unstable for a second
critical value $A^*_{nu,2}$ $(=0.529\,159 \cdots)$ through a
period-doubling bifurcation, giving rise to the birth of a symmetric
orbit of period $2$ denoted by the solid line.

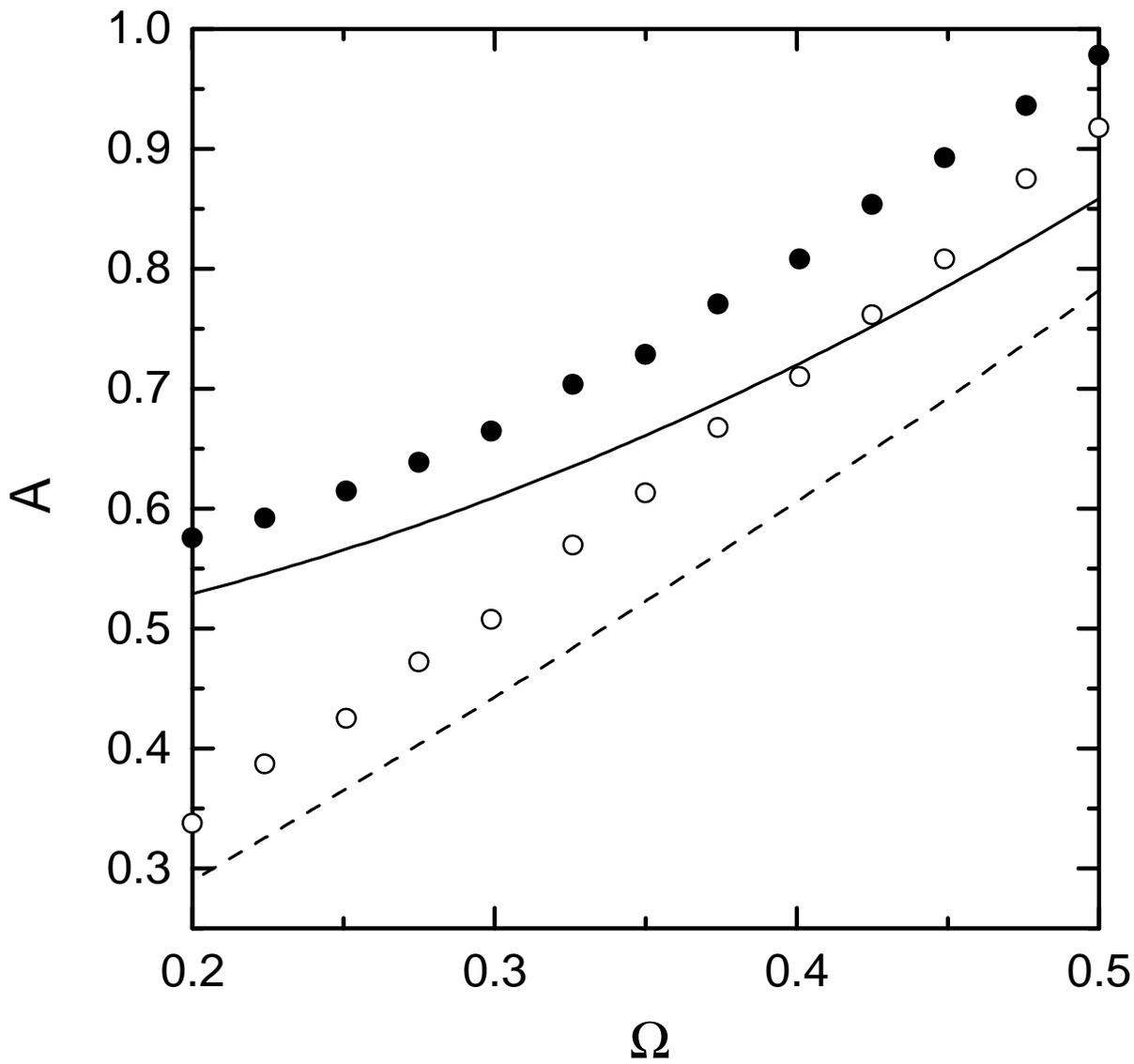
\begin{figure}
\caption{Stability  diagram of  the inverted SP in the range of
         $0.2 \leq \Omega \leq 0.5$. The experimental data for the
         1st and 2nd critical values, $A^*_{exp,1}$ and
         $A^*_{exp,2}$, are represented by the open and solid
         circles, respectively. The lower and upper stability
         boundaries, numerically computed using the Floquet theory
         and denoted by the dashed and solid lines,
         are the symmetry-breaking pitchfork and
         period-doubling bifurcation lines, respectively.
         }
\label{fig:ISD}
\end{figure}

We also carry out the above experiments for many other values of
$\Omega$, and thus measure the 1st and 2nd critical values,
$A^*_{exp,1}$'s and $A^*_{exp,2}$'s. Figure \ref{fig:ISD} shows the
stability diagram of the inverted SP. The experimental data for
$A^*_{exp,1}$ and $A^*_{exp,2}$ are represented by the open and
solid circles, respectively. The inverted SP is observed to become
stable in the interval between $A^*_{exp,1}$ and
$A^*_{exp,2}$. Note also that the width of this interval becomes
smaller as $\Omega$ is increased. Hence the stabilization of the
inverted SP can be more easily observed for small values of $\Omega$,
compared to the cases of high $\Omega$-values.
For a comparison with the experimental results,
numerical data obtained using the Floquet theory are also given in
Fig.~\ref{fig:ISD}. The lower stability boundary $A^*_{nu,1}$
denoted by the dashed line is a subcritical pitchfork bifurcation
line, while the upper stability boundary $A^*_{nu,2}$ denoted by the
solid line is a period-doubling bifurcation line.
We note that the agreement between the experimental and
numerical results becomes better as $\Omega$ is decreased.

\section{Summary}
\label{sec:SUM}

Bifurcations of the normal and inverted SP's in the parametrically
forced MP are experimentally studied by varying the two parameters
$\Omega$ and $A$. As $A$ is increased above a critical value,
the normal SP is observed to become unstable either by a
period-doubling bifurcation or by a symmetry-breaking pitchfork
bifurcation, depending on the values of $\Omega$.
In the 1st range $I$ of $\Omega$ (i.e., $0.2 \leq \Omega \leq 0.5$),
a new stable symmetric orbit with period $2$ appears via
period-doubling bifurcation, while a conjugate pair of new stable
asymmetric orbits with period $1$ is born via symmetry-breaking
pitchfork bifurcation in the 2nd range $II$ of $\Omega$ (i.e.,
$0.8 \leq \Omega \leq 1.025$). However, in contrast to this normal SP,
the inverted SP is observed to become stable when $A$ is increased
above a 1st critical value $A^*_1$ by a subcritical pitchfork
bifurcation. Unfortunately a pair of asymmetric orbits of period $1$
born for this subcritical case cannot be experimentally
observed, because they are unstable ones. As $A$ is further
increased, the stabilized inverted SP is also observed to lose its
stability for a 2nd critical value $A^*_2$ by a period-doubling
bifurcation, giving rise to the birth of a new stable symmetric
period-doubled orbit. Thus the inverted SP becomes stable in the
interval between $A^*_1$ and $A^*_2$. When all
of these experimental results for the two SP's are compared with
the numerical results obtained using the Floquet theory, they seem to
agree well.

\acknowledgements

This work was supported by the Basic Science Research Institute
Program, Ministry of Education, Korea, Project No. BSRI-96-2401.

\begin{references}
\bibitem{BE} By decreasing the value of $B_{DC}$ from some positive
value, we first get a data set of $\omega_0(B_{DC})$ for several
values of $B_{DC}$ through the fitting procedure described in the
4th paragraph in Sec.~\ref{sec:ES}. We can then obtain the values of
$B_{E,n}$ and
$m/I$ by fitting the experimental data set of $\omega_0(B_{DC})$ to
the equation, $\omega_0 = \sqrt{ {m \over I} (B_{DC} + B_{E,n}) }$.
\end{references}
\end{document}